\pdfoutput=1
\documentclass[twocolumn,prl,amsmath,letterpaper,floatfix,showpacs]{revtex4}

\usepackage{epsfig}
\usepackage{graphicx}
\usepackage{amsmath}
\usepackage{amssymb}
\usepackage{bm}

\def\prl#1#2#3{Phys.~Rev.~Lett.~{\bf #1},\ #2\ (#3)}

\newcommand{\Xs} {{X^1\Sigma^+}}
\newcommand{\Xt} {{a^3\Sigma^+}}
\newcommand{\As} {{A^1\Sigma}}
\newcommand{\At} {{b^3\Pi}}

\newcommand{\1}{{|1\rangle}}
\newcommand{\2}{{|2\rangle}}
\newcommand{\3}{{|3\rangle}}

\renewcommand{\a}    {{\alpha}}

\begin{document}

\title{Piecewise adiabatic population transfer in a molecule via a wave packet}

\date{\today}
\author{Evgeny A.~Shapiro$^1$, Avi Pe'er$^{4}$, Jun Ye$^4$, Moshe Shapiro$^{1,2,3}$}
\affiliation{ %
Departments of $^1$Chemistry and $^2$Physics, The University of British Columbia, Vancouver, V6T 1Z2, Canada,
\\ $^3$Department of Chemical Physics, The Weizmann Institute, Rehovot, 76100, Israel,
\\ $^4$JILA, National Institute of Standards and Technology and University of Colorado, Boulder, CO 80309-0440, USA.}

\begin{abstract}
{ We propose a class of schemes for robust population transfer
between quantum states that utilize trains of coherent pulses and
represent a generalized  adiabatic passage via a wave packet. We
study piecewise Stimulated Raman Adiabatic Passage with
pulse-to-pulse amplitude variation, and piecewise chirped Raman
passage with pulse-to-pulse phase variation, implemented with an
optical frequency comb. In the context of production of ultracold
ground-state molecules, we show that with almost no knowledge of
the excited potential, robust high-efficiency transfer is
possible. }
\end{abstract}

\pacs{42.50.Hz, 82.50.Nd, 82.53.Kp, 34.50.Cx}



\maketitle

With the advent of technologies that deal with quantum properties
of matter, there is a need for methods of controlling quantum
systems that are, on one hand, robust and conceptually simple, and
on the other, flexible and applicable far beyond few-level
arrangements. To address this need, our goal in this work is to
utilize the intuition and control recipes of simple schemes for
robust transfer of quantum population based on Adiabatic Passage
(AP), and apply them to many-level quantum systems with complex
dynamics. The key components of the method are
coherent accumulation of amplitudes
 by a train of phase coherent optical pulses, and interference of quantum pathways in multi-state population
transfer.

We test our ideas, simulating dynamics in KRb molecules, which is
of special interest because of the prospect for creating deeply
bound ultracold polar molecules starting from loosely bound
Feshbach states \cite{FeshbachMolecules,DebbieRecentPRL}. As
translationally and vibrationally cold polar molecules are a
prerequisite for many applications \cite{ultracold-review}, their
robust high-yield creation is a forefront goal.

{Coherent accumulation has been studied in the context of
perturbative control \cite{Nelson-ChemRev94} and later on in
precise spectroscopy using frequency combs \cite{DFCS}. Its use
for \emph{complete} population transfers was studied so far
twice.} One was the analysis within the 3 level model of piecewise
adiabatic passage (PAP) 
\cite{PAP-1}; the other study demonstrated population transfer
through a wave packet with a coherent train of pump-dump pulses
\cite{Avi_PA}. Here, we unite the two as special cases within one
general framework, where the combination of adiabatic transfer
concepts with a coherent train excitation (optical frequency comb)
preserves the robustness of adiabatic transfer, but is applicable to
composite quantum states.

The concept of piecewise adiabaticity is elucidated with two main
examples. In piecewise stimulated Raman adiabatic passage
(STIRAP), robust transfer is achieved through a slow variation of
the {intensity envelope} of the driving pulse trains. In the
other, which we term ``piecewise chirped Raman passage'' (CRP),
robustness is obtained by slow variation of the excitation phase.
We demonstrate how both examples can be extended to the case of
the intermediate state being a wave packet, i.e. a multiplicity of
intermediate states.
In the spectral domain, the
transfer is seen as a constructive interference of several AP pathways \cite{CCAP-review,SBook}. Last, we discuss
the case of transfer without detailed knowledge of the intermediate dynamics. We show that by scanning the comb
parameters of a train of {unshaped pulse-pairs ({repetition rate and intra-pair time separation})} it is possible
to achieve high efficiency transfer. The intermediate dynamics can then be studied using a two-dimensional
mapping of the transfer efficiency as a function of the comb parameters.

A traditional AP transfers population from an initial to a target
state by dressing the quantum system in slowly changing fields. If
one of the time-dependent field-dressed eigenstates coincides with
the initial state at the beginning of the process, and with the
target at the end \cite{RiceBook,Oreg-AP-PRA84, stirap,SBook}, a
complete and robust transfer is obtained as long as the dressing
fields change adiabatically. Accordingly, the prescription of PAP
\cite{PAP-1} can be summarized as: For reference, consider any
traditional AP, driven by slowly varying fields. Then, break the
reference AP into a set of time intervals $\tau_n$, so short that
only a small fraction of population is transferred between the
eigenstates during each interval. Within each interval, the driving
fields can be replaced by {\it any other} field amplitude shape, as
long as the integral action of each of the fields over $\tau_n$
remains the same. For example, the smooth reference field can be
replaced by a train of mutually coherent pulses. Last, vary the
inter-pulse 'silent' time of the system's free evolution.
Controlling the silent, free evolution durations and the phase of
each of the driving fields allows control over the arising
Ramsey-type interference picture.

\begin{figure}
\centering
    \includegraphics[width=1.0\columnwidth]{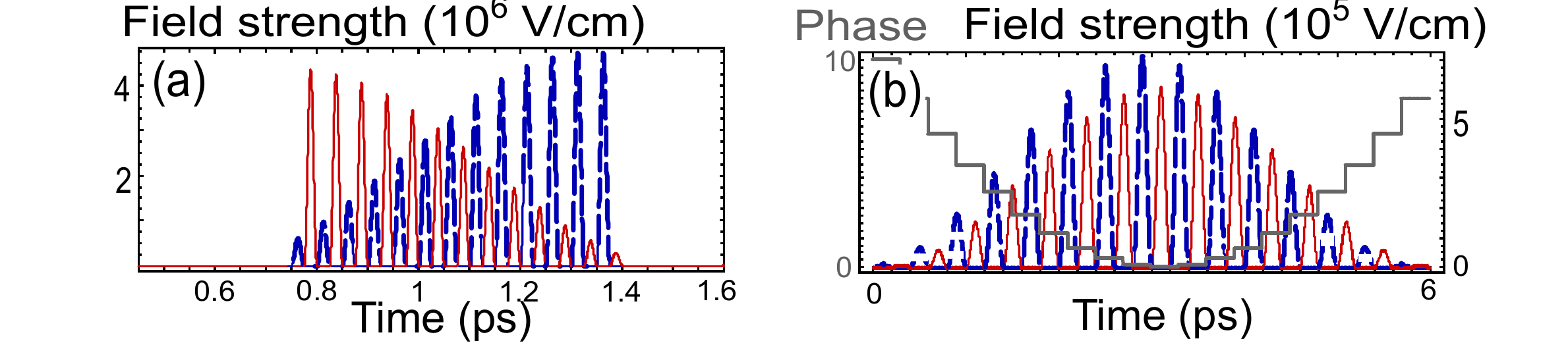}
    \caption{(Color online). Sequences of femtosecond pulses in piecewise STIRAP (a) and piecewise
    CRP (b) between states $\1=3S$, $\2=4P$, and $\3=5S$ of Na \cite{Na-data}. Dashed blue:  pump field
    envelope; solid red: dump field envelope; gray: the carrier phase in piecewise CRP with $\a_P=\a_D=0.2$.
 }
     \label{FigNa}
\end{figure}

Ref.~\cite{PAP-1} introduced piecewise versions of STIRAP
\cite{Oreg-AP-PRA84, stirap}, which transfers population in a
three-level system from state $|1\rangle$ into state $|3\rangle$ via
the intermediate $|2\rangle$. {In the key example}, the field was
given as a train of femtosecond ``pump'' and ``dump'' pulse pairs,
resonantly coupling the eigenstates $\1$ and $\2$ (pump), and the
eigenstates $\2$ and $\3$ (dump). The pulses -- ``kicks'' -- did not
overlap in time and the pulse train envelopes were slowly varied to
achieve PAP, as shown in Fig. \ref{FigNa}(a).

Another conventional AP in a three-level system is CRP
\cite{Oreg-AP-PRA84}. The pump and dump pulses share a similar
smooth temporal intensity profile; however, both pulses are
frequency chirped, so that their phases are $ \phi_{P,D}(t)=
\omega_{P,D}t + \alpha^{(\omega)}_{P,D}(t-t_0)^2/2 $. Like STIRAP,
CRP transfers population from state $\1$ into $\3$ in a robust
manner; unlike STIRAP, CRP creates transient population in state
$\2$. {For the piecewise version, all pulses are exactly on
resonance. The effect of chirp is mimicked by varying the carrier
phase of each pulse
$\phi_{P,D}^{(0)}(n)=\phi_{P,D}(n,t)-\omega_{P,D}t$ quadratically
with the pulse number $n$:}
\begin{equation}
\phi_{P,D}^{(0)}(n)= \a_{P,D}(n-n_0)^2/2 %
\label{QuadraticPhase}\end{equation}%
as shown in Fig.~\ref{FigNa}(b). Pulse trains with such a piecewise chirp demonstrated piecewise adiabatic
following in a two-state system \cite{valery-PAF}. Here we observe that the piecewise CRP in a three-level system
is robust with respect to the field strengths, the number of pulse pairs, the values and signs of $\a_P$ and
$\a_D$, and to additional delays between the pump and dump
pulse trains. 

In both foregoing examples, {the details of the single pulse
dynamics are not important, only the values of the basis state
amplitudes after each $\tau_n$ matter.} Based on this observation,
below we extend the conceptual scheme of PAP, generalizing the
eigenstate $\2$ to a multitude of states, forming a wave packet.
This wave packet undergoes complex dynamics between the pairs of
pump and dump kicks, but if the pulse repetition time coincides
with a revival, the wave packet (almost) returns to its original
state by the time the next pair of kicks arrives. The kicks, in
turn, become rather complex operations implemented via an
interplay of the shaped femtosecond pulses with the intermediate
wave packet dynamics. We expect that, as long as the action of the
kicks on the basis states $\1$, $\2$, and $\3$ mimics that of the
dressing fields in the reference AP coarse-grained over $\tau_n$,
the enforced population transfers in both schemes is insensitive
to the parameters of the driving fields. Further,
high fidelity of the revivals turns out to be unnecessary, as we
demonstrate later with unshaped pulses.

\begin{figure}
\centering
   \includegraphics[width=0.85\columnwidth]{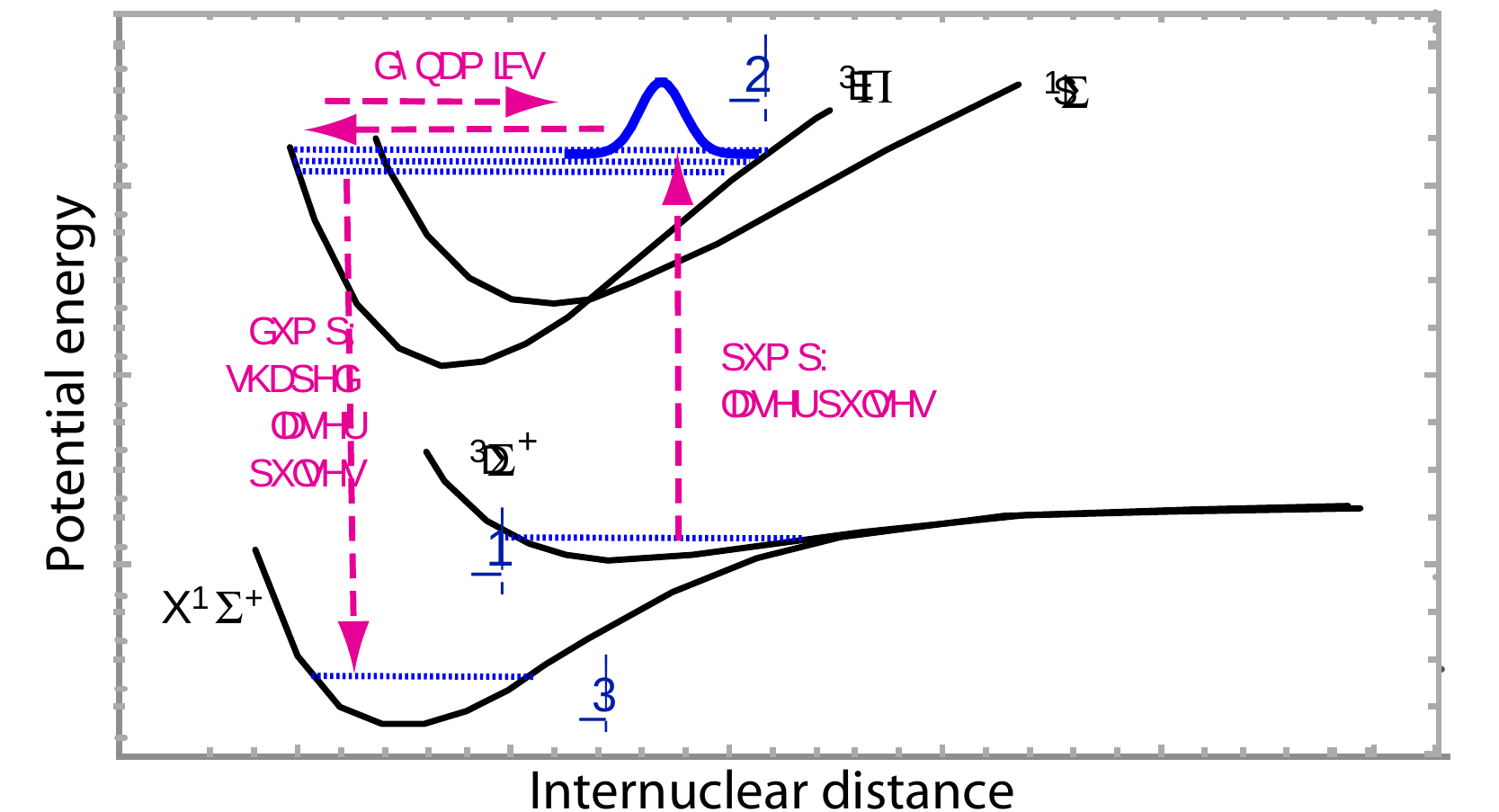}
    \caption{(Color online). KRb potentials, basis states, and PAP arrangements discussed in the text.}
 \label{FigPassages}\end{figure}

Below we study this general idea simulating creation of deeply
bound $\Xs$ KRb molecules from a $\Xt$ state.
The molecular potentials used in the simulations are shown
schematically in Fig.~\ref{FigPassages}. Eigenenergies and
transition strengths are calculated using the algorithm FDEXTR
\cite{fdextr} with the data input from
Refs.~\cite{KRb-pots,kotochigova-KRbdipoles}. The excited
electronic state is modeled as the LS-coupled $\As$ and $\At$
potentials.
{Although the model is too simple to precisely describe molecular
dynamics on nano-microsecond time scales, it retains the main
complexity of the problem due to the coupled singlet-triplet
evolution.} We choose $\1$ to be the vibrational state $v=5$ ($E=
-157$ cm$^{-1}$) of the $\Xt$ potential (Fig.~\ref{FigPassages}). To
trace the fidelity of the method, we also consider the amplitudes of
vibrational states around the input state ($v=0,1,...,12$ of the
$\Xt$ potential). The wave packet $|2\rangle$ is composed of up to
20 vibrational states of the LS coupled $\As$-$\At$ potentials with
the energies $E = 11000 - 11400$ cm$^{-1}$. State $\3$ is $v=22$
($E= -2490$ cm$^{-1}$) of the ground $\Xs$ potential; the
neighboring $\Xs$ states $v=16,...,28$ are included in the
simulations as well.

\begin{figure} \centering
    \includegraphics[width=0.99\columnwidth]{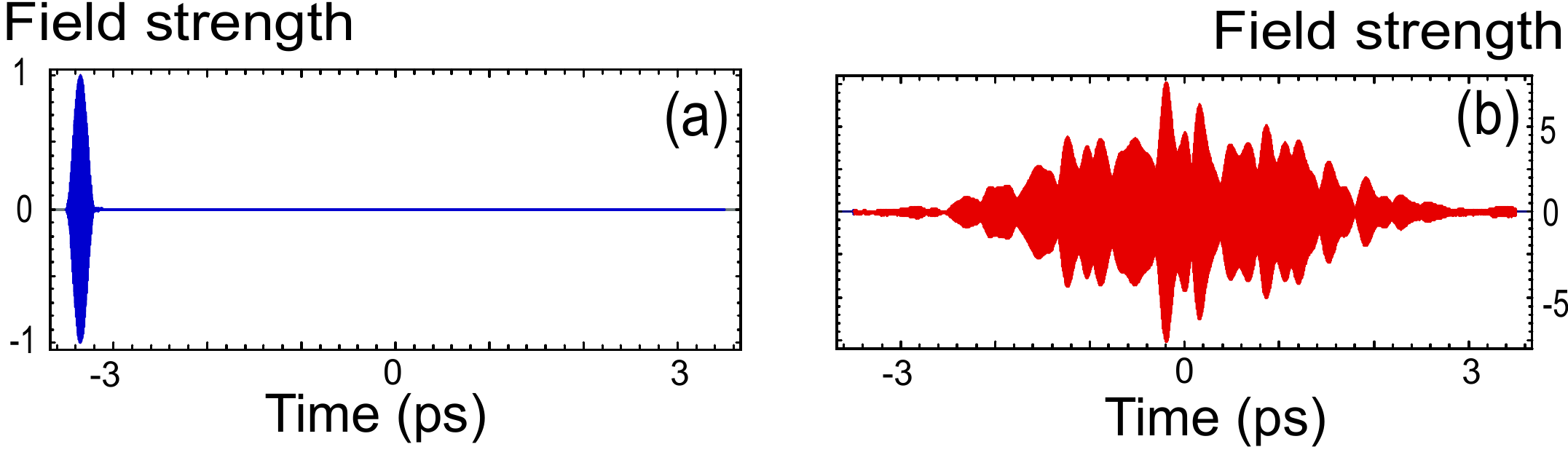}
    \caption{(Color online) Time-dependent field of the pump (a), and dump
    (b) pulses. The field in each pulse is normalized relative to the maximum of the pump field.
       }     \label{FigPulses}
\end{figure}

The system, initially in state $\1$, is driven by a series of
mutually coherent pump and dump femtosecond pulses. A single pump
pulse ($110$ fs FWHM in intensity, $\sin^2\alpha t$-shaped, as
shown in Fig.~\ref{FigPulses}(a)), excites an $\As$-$\At$ wave
packet, which oscillates on the coupled $\As$-$\At$ potential
surfaces with the singlet and triplet vibrational periods of the
order of $800-900$ps. After several vibrations the dynamics become
irregular. {However, at certain times, which correspond to
two-surface revivals, the wave packet} rephases with a fidelity
reaching $\sim0.9$. In our simulations, high-fidelity revivals
were observed, roughly, every 80 ps; the particular revival time
is sensitive to the spectroscopic data used as input. Following
its excitation, the wave packet can be dumped into the single
vibrational level $\3$. The shape of the femtosecond dump pulse is
found by requiring that the wave packet before dumping overlaps
well with the wave packet that would have been excited by the
time-reversed dump, in a manner similar to the prescription of
Ref.~\cite{Avi_PA}. The thus-found dump field is shown in
Fig.~\ref{FigPulses}(b).

\begin{figure} \centering
    \includegraphics[width=0.99\columnwidth]{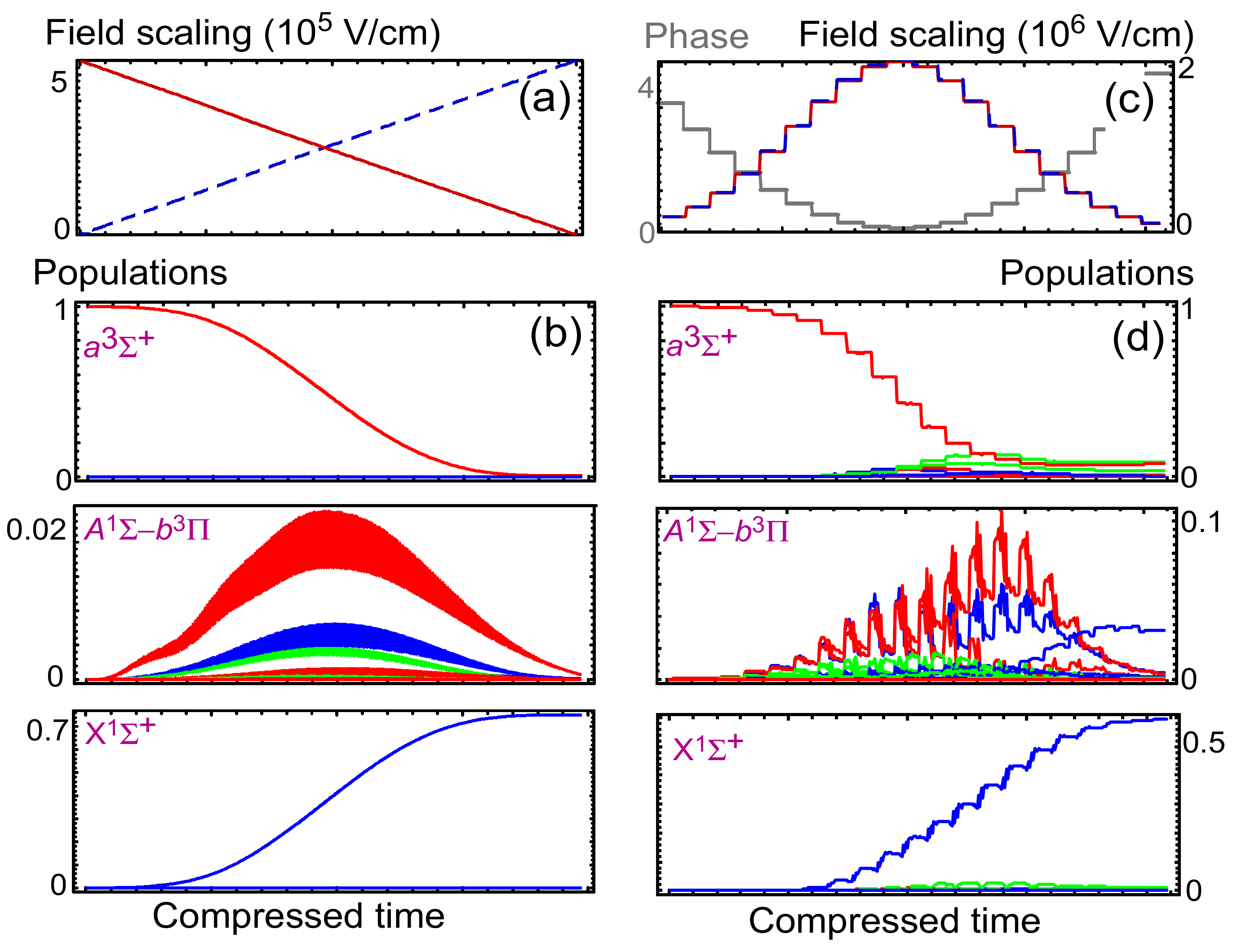}
    \caption{(Color online) Dynamics of the STIRAP-like (a,b) and CRP-like (c,d) PAP in KRb. (a,c): Envelopes of the pump (dashed blue) and dump (red) pulses.
    Each point shows the factor by which the field shown in Fig.~\ref{FigPulses} is scaled. The carrier phase $\phi^{(0)}$
    in the piecewise CRP, same for the pump and dump, is shown by the gray staircase in
    (c). (b,d): Population dynamics of all the levels in the system.
    Time is given at the ``compressed'' set of points: output is shown only for {the times} when the field
    is on.
       }     \label{FigPAPs}
\end{figure}

The simulation results for both piecewise STIRAP and piecewise CRP
are presented in Fig.~\ref{FigPAPs}. For STIRAP, shown in
Fig.~\ref{FigPAPs}(a,b), the field consists of 200 pairs of pump and
dump pulses, all shaped as in Fig.~\ref{FigPulses}. The trains
envelopes vary linearly on the pulse number (Fig.~\ref{FigPAPs}(a)).
The inter-pair separation is 1310.59 ps, which is close to a revival
time of the wave packet $\2$. The carrier frequencies are chosen to
match the Raman condition $\omega_{Raman}=\omega_P-\omega_D$. The
carrier optical phase is kept constant throughout each pulse train
by choosing the carrier frequency to coincide with a comb tooth
\cite{PAP-1}:
 \begin{equation}\label{frequencies}
\omega_{P,D}=2\pi (N_{P,D}f_{rep}+f_{0\,P,D}) \ ,
\end{equation} %
where $f_{rep}$ is the common repetition rate and $f_{0\,P,D}$ is
the carrier-envelope offset frequency in the pump (dump) train.
The process is interferometrically sensitive to the inter-pulse
delay ($f_{rep}$) and requires phase locking of the pump and dump
combs; however, phase stabilization of frequency combs is now a
standard technique \cite{DFCS}.

If the spontaneous decay of the $\As$-$\At$ states is neglected,
the calculation predicts transferring 92\% of population from $\1$
into $\3$. Introducing a 15-ns $\As$-$\At$ decay time reduces the
transfer efficiency to 75\%. The maximum combined transient
population of all the $\As$-$\At$ levels reaches $\sim$0.04.
The
transfer efficiency is found to be rather insensitive to the field
strength profiles in the pulse trains, and varies within about
15\% if the number of pulses is varied between 35 and 300.

Figure~\ref{FigPAPs}(c,d) shows an example of CRP-like PAP. The
field consists of 20 pairs of pump and dump pulses. The pump and
dump pulse train envelopes are Gaussian (Fig.~\ref{FigPAPs}(c)).
The inter-pair separation is 1309.93 ps. The central carrier phase
in each field evolves with the pulse number according to
Eq.(\ref{QuadraticPhase}), with $\a_P=\a_D=0.09$. The calculation
with decaying $\As$-$\At$ levels predicts transferring $60\%$ of
the initial population into the target state; neglecting the decay
raises this number to $65\%$. Approximately 2\% of the population
ends up distributed among eigenstates other than $\3$ of the
target $\Xs$ manifold; 22\% of the initial population are
transferred to neighboring states of $\1$ in the $\Xt$ manifold at
the end of the process. Unlike in the simple 3-level case,
CRP-like transfer in the molecule is found to be significantly
more sensitive than the STIRAP-like transfer to the number of
pulses and the intensity profiles of the pump and dump pulse
trains. We note that the required train envelope modulations
(either in amplitude or phase) can be readily generated
experimentally by present modulators for standard 1-10 ns
repetition-period pulse sources.

The dependence of the transfer efficiency on $\Delta T$ in both STIRAP-like and CRP-like PAP is irregular due to
the complex dynamics of the $\As$-$\At$ wave packet. Viewed from the spectral perspective, each of the
$\As$-$\At$ eigenstates can serve as an intermediate level for a separate AP pathway. Reminiscent of the
traditional coherent control via a wave packet \cite{SBook,Albert-WPMetr-98}, the AP pathways interfere to
provide the population transfer into the target state. Thus the frequency comb source drives a coherently
controlled AP process \cite{CCAP-review, DFCS} via the set of intermediate levels. Matching the repetition time
and carrier phase to the wave packet revival time and phase ensures that all the possible AP pathways
participate; {spectral shaping of the pump and dump pulses sets constructive interference among the pathways.}

\begin{figure} \centering
   \includegraphics[width=0.99\columnwidth]{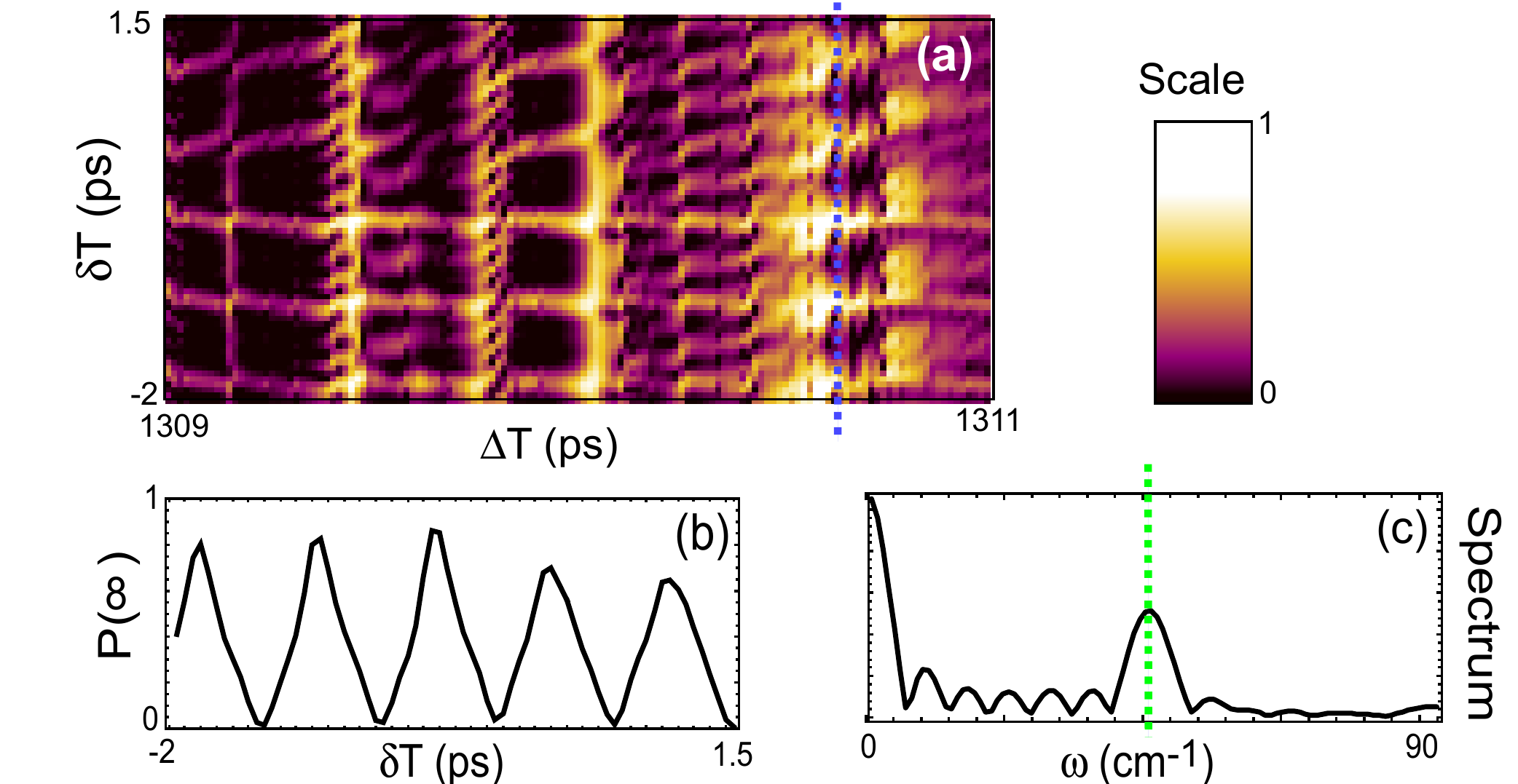}
\caption{(Color online)(a): An efficiency density plot for the STIRAP-like PAP with unshaped laser pulses in
dependence on the inter-pair delay $\Delta T$ and the intra-pair delay $\delta T$. Positive $\delta T$ indicates
that the pump comes {after} the dump; i.e. population remains in the excited wave packet until the next dump
comes (suffering more spontaneous emission losses). (b): Vertical at $\Delta T = 1310.62$ ps (blue dotted line in
(a)). (c): Fourier transform of (b). The green dotted line marks the beat frequency between $E_{2}^{(1)}$ and
$E_{2}^{(2)}$.}
       \label{FigUnshaped}
\end{figure}

Tailoring the optimal pulses requires detailed knowledge of the
excited state structure, which is not always available. But what
happens if the pulses are not shaped at all?
Fig.~\ref{FigUnshaped}(a) investigates the total efficiency of the
STIRAP-like transfer under {\it both} pump and dump trains
consisting of 50 $\sin^2\a t$-shaped pulses, with respect to the
inter-pair delay $\Delta T=1/f_{rep}$ and intra-pair delay $\delta
T$ between the pump and dump pulses. Raman resonance is maintained
throughout the $f_{rep}$ scan by varying the carrier-envelope
frequency difference according to
$f_{0\,P}-f_{0\,D}=(\omega_{Raman}/2\pi)-Nf_{rep}$. The plot shows
a remarkable landscape of vertical and horizontal features, where
some peaks reach 80\% transfer efficiency.
Note that no population is transferred to any of the $\Xt$ or $\Xs$
states, except $\1$ and $\3$, or to any of the intermediate states.
In the spectral domain, either one state guides one of the
interfering AP pathways, or it is simply missed by the narrow
spectral selectivity of the frequency combs. In the time domain,
destructive Ramsey-type interference prevents transitions to
undesired levels.

To elucidate some of the structure observed in
Fig.~\ref{FigUnshaped}, consider $\delta T$ expressed in frequency
as a linear spectral phase of the dump comb with respect to the
pump comb. Accordingly, the phase of a specific Raman path at
detuning $\Omega_i$ from the carrier frequency is
$\varphi(\Omega_i)= -\Omega_i \delta T+\phi_{d}(\Omega_i)$, where
$\phi_{d}(\Omega_i)$ is the relative phase of the dipoles
associated with this path. The vertical lines in the image
correspond therefore to a $\Delta T$ value where one or several
teeth are in one-photon resonance with Raman paths. Changing
$\delta T$ at this $\Delta T$ reveals interference between these
paths. A Fourier transform with respect to $\delta T$ reveals the
spacing between the intermediate levels, as shown in
Fig.~\ref{FigUnshaped}(b,c), where mostly two Raman paths, via
$E_{2}^{(1)}=11145$ cm$^{-1}$ and $E_{2}^{(2)}=11190$ cm$^{-1}$,
contribute, and the Fourier transform of the transfer efficiency
peaks at the beat frequency between $E_{2}^{(1)}$ and
$E_{2}^{(2)}$. Horizontal lines, on the other hand, occur when
$\delta T$ hits a special value, where
$\varphi_{d}(\Omega_i)\approx \Omega_i \delta T$ for the
participating Raman transitions; i.e., $\delta T$ is matched to
the vibrational dynamics. In this case the transfer peaks even if
there is no coherent accumulation in the intermediate levels, as
was observed already in \cite{Avi_PA}. Since the match to the
dynamics is only approximate, the lines are not perfectly
horizontal. In the best case scenario every dump pulse transfers
into $\3$ exactly all the population excited to $\2$ from $\1$ by
the preceding pump \cite{Avi_PA}.

With some preliminary knowledge of the initial and target state's
energies, the range of frequencies and the field strengths needed
(but not of the exact values of intermediate energies and
transition dipoles), one can take trains of unshaped pump and dump
pulses, and scan $\Delta T$ and $\delta T$  to find efficient PAP
transfers. In an analogy with 2D Fourier spectroscopy
\cite{2D-FT-spectro}, a theoretical analysis of the 2D scan of the
transfer efficiency may be able to provide the full spectroscopic
information. Further, one can experimentally optimize the pulse
shapes in order to maximize the transfer fidelity.

We thank V. Milner, I. Thanopulos, C. Koch, and S. Kotochigova for
discussions and consultations. The work at JILA was supported by
NIST and NSF.

\end{document}